\documentclass[twocolumn,prd,showpacs]{revtex4}
\usepackage{graphicx}% Include figure files
\usepackage{dcolumn}% Align table columns on decimal point
\usepackage{bm}% bold math

\begin{document}

\title{End-point of the {\it rp} process and periodic 
gravitational wave emission}

\author{P. B. Jones}

\affiliation{Department of Physics, Denys Wilkinson Building,\\
University of Oxford, Keble Road, Oxford OX1 3RH, England}

\date{\today}% It is always \today, today,
             %  but any date may be explicitly specified

\begin{abstract}
The general end-point of the {\it rp} process in rapidly accreting
neutron stars is believed to be a surface distribution of matter whose
nuclear composition may depend on position.  
Its evolution during compression beyond the neutron-drip threshold density
$\rho_{nd}$ is determined by the presence of nuclear formation
enthalpy minima at the proton closed shells. At $\rho_{nd}$, a sequence of
weak interactions with capture or emission of neutron pairs rapidly
transform nuclei to the most accessible proton closed shell.  Therefore,
angular asymmetries in nuclear composition present in accreted matter at
$\rho_{nd}$ are preserved during further compression toward densities
$\sim 10^{14}$ g cm$^{-3}$ provided transition rates between closed
shells are negligible. Although it has been confirmed that this condition
is satisfied for predicted internal temperatures and for the formation
enthalpy distribution used in this work, it would not be so if the true
enthalpy differences between maxima and minima in the distribution were
a factor of two smaller.  For this reason, it does not appear possible
to assert with any confidence that position-dependent surface composition
can lead to significant angle-dependence of the equation of state and to
potentially observable gravitational radiation. The effect of non-radial
internal temperature gradients on angle-dependency of the equation of
state is also not quantifiable.

\end{abstract}

\pacs{97.60.Jd, 97.80.Jp, 95.55.Ym, 26.60.+c}% PACS, the Physics and Astronomy
                             % Classification Scheme.
%\keywords{Suggested keywords}%Use showkeys class option if keyword
                              %display desired
\maketitle

\section{\label{I}Introduction}

Publication of the first data from the GEO 600 and LIGO interferometric
gravitational wave detectors \cite{abb04} is a notable event in the study
of neutron stars. Upper limits for periodic emission were obtained,
initially for PSR J1939+2134, and then for a further
27 isolated pulsars.  The current rate of improvement in 
sensitivity makes the detection of signals from these and similar
galactic sources an interesting future possibility.   Gravitational
wave emission occurs if the rotating neutron star has mass-quadrupole
moment tensor \cite{ll62} components that are time-dependent
in the observer frame.  Magnetic stress is an obvious
possible source of such deformation, particularly if the internal
fields are one or more orders of magnitude larger than those inferred
at the surface. A brief recent survey of this topic has been given by
Cutler \cite{cut02}.  Furthermore, it has been recognized, within the
last ten years, that high rates of mass accretion in low-mass X-ray
binary systems may also lead to neutron star deformation and gravitational
wave emission \cite{bil98,ucb00,mp01}.

Burning of accreted hydrogen
and helium in the neutron star atmosphere has been the subject of a very
large number of published papers.  Ref. \cite{bb98,sbcw99,sabb01,khkf04} 
give results from more recent calculations of the distributions
of nuclear mass number $A$ and charge $Z$ which are formed by rapid
proton capture (the {\it rp} process) during stable or unstable burning.
The predicted distributions
are functions of the thermodynamic conditions existing during the
{\it rp} processes, indicating the possibility of non-trivial differences
in atmospheric composition between, for example, the magnetic polar
and equatorial regions of the star.  
The accretion rates considered can be extremely large 
($\sim 10^{-8}$ M$_{\odot}$ yr$^{-1}$) and the consequent inward flow
of matter is so rapid
that any dependence of atmospheric composition on the angular coordinates
$\theta,\phi$ which define position on the neutron-star surface must
certainly be transferred
to the liquid ocean which is present above the solid phase of the crust.
Unless there is very strong convective instability it will
survive there for times $\sim 1$ yr before entering
the solid phase at a matter density $\rho\sim 10^{9}$ g cm$^{-3}$.
(Fig. 1 of Ref. \cite{bb98} gives a useful
summary of relevant orders of magnitude for surface layer properties.)
An element of accreted matter moving inward is compressed and undergoes
a sequence of nuclear
transitions as the density $\rho$ increases.  Finally, there may be
changes of phase before entry to the liquid core of the neutron star.     
Sato \cite{sat79} first recognized that the composition of this
accreted matter, as a function of $\rho$, would be quite different
from that of primordial neutron-star matter.  Detailed calculations
of the sequence of transitions starting from $(A,Z)=(56,26)$ at
$\rho=10^{8}$ g cm$^{-3}$ have been made by Haensel and Zdunik
\cite{hz90} who also repeated this work \cite{hz03} for the initial
condition $(A,Z)=(106,46)$ following publication of Ref. \cite{sabb01}.
Below the neutron-drip threshold $\rho_{nd}$, matter consists of nuclei
embedded in a relativistic electron gas.  The transitions in this region
are a sequence of electron capture reactions which reduce $Z$ at constant
$A$.  Thus any position-dependent differences in nuclear mass number
present in the atmosphere and surface ocean are preserved at $\rho<\rho_{nd}$
so that the equation of state may have, in principle, a small
$\theta,\phi$-dependent component \cite{bil98,ucb00}. Although this
region contains only $\sim 10^{-5}$ M$_{\odot}$, the neutron-star
mass-quadrupole tensor components generated by it are not necessarily
negligible.  Even if surface ocean composition were uniform, a
non-radial temperature gradient $\nabla _{\theta,\phi}T\neq 0$ present
within the solid crust
changes the densities at which the electron capture transitions occur
and so leads, in principle, to a small angle-dependent component in the
equation of state \cite{bil98,ucb00}.

The inner part of the solid crust at $\rho>\rho_{nd}$ is more massive
$(\sim 10^{-2}$ M$_{\odot})$ and therefore the investigation of possible
angle-dependency in composition and equation of state in this region is a
very important problem.  It has been considered recently by Ushomirsky,
Cutler and Bildsten \cite{ucb00} in calculations based on the work of
Sato \cite{sat79} and of Ref. \cite{hz90}. The analysis of the evolution
of nuclear charge at $\rho > \rho_{nd}$
given in the present paper is in substantial disagreement with
Ref. \cite{ucb00} because these authors assume,
following Sato \cite{sat79} and Ref. \cite{hz90},
that accreted nuclei at $\rho > \rho_{nd}$ follow an evolutionary
path of decreasing $Z$, produced by successive electron captures,
until transition rates for pycnonuclear fusion reactions become
appreciable at $Z\approx 10$.  
Calculations of nuclear formation enthalpies \cite{jo04} at
$\rho > \rho_{nd}$, though necessarily very approximate, show that
local minima exist at the proton closed shells with
$Z=20, 28, 34, 40, 50$.  Our analysis is that the presence of these
minima determines the evolution of accreted matter during compression
from the neutron-drip threshold until $\rho$
approaches the region of phase transitions to possible lower-dimensional
structures, or the liquid core of the star.  The internal
temperature and the nature of the formation enthalpy distribution are
the crucial factors that determine whether or not
the system reaches weak-interaction equilibrium (homogeneity in $Z$) 
in the $10^{6}-10^{7}$ yr interval of high accretion rates.
Sec. IIA contains a brief description of the $Z$-dependence of
the nuclear formation enthalpy which is the basis of the present paper.
A system in which more than one closed shell is populated must
eventually become unstable, during compression to
higher densities, against quantum-mechanical tunnelling of protons
between adjacent nuclei, even if weak-interaction transition rates
are negligibly small (Sec. IIB).
We also consider the inward movement of the accreted matter and show
in Sec. IIIA that the stress-response of the solid cannot, without
qualification, be
described as elastic. Calculation of the quadrupole moment produced
by the angle-dependent equation of state is therefore not a simple
problem and for this reason the solid crust is represented as a
one-dimensional system in Sec. IIIB.  A brief comparison is given
with the one-dimensional treatment considered in Ref. \cite{ucb00}.

Whether or not periodic signals are seen by interferometric detectors,
it is
important to have some understanding of the ways in which they might
be generated.  Magnetic stress is the more simple source, although
questions such as the form of the field distribution and the nature
of proton superconductivity present themselves.  The main conclusion
of this paper is that our present understanding of the relevant aspects
of nuclear structure in the neutron-drip region is such that predictions
of gravitational wave emission arising from compositional asymmetries
cannot be made with confidence.  But the significant factors are
considered in Sec. IV.

\section{\label{II}The structure of the neutron-drip solid}
\subsection{\label{A}Shell effects above the neutron-drip threshold}

The canonical view of the solid phase has been
that strong and weak-interaction equilibria are maintained during
cooling so that a Coulomb {\it bcc} lattice of nuclei is formed on
solidification, neutralized by a relativistic electron gas.  The matter
density at the neutron-drip threshold has some dependence on the details
of the equation of state but is $\rho_{nd}\approx 4-8 \times 10^{11}$
g cm$^{-3}$.  At $\rho>\rho_{nd}$, the neutron chemical potential
exceeds its rest energy and the nuclei are in equilibrium with both the
electrons and a
continuous neutron medium, superfluid at temperatures below $T
\sim 10^{9}$ K.  The early and classic paper of Negele and Vautherin
\cite{nv73}, involving microscopic calculations of the single-particle
states for neutrons and protons inside a Wigner-Seitz cell, gave the
structure of nuclei within the neutron medium. The energy differences
per nucleon, at fixed baryon density, between lattices with different
proton closed shells, for example, $Z= 40, 50$ were found to be very
small, indicating that the standard assumption of a homogeneous
{\it bcc} lattice might be questionable.  Much later, calculations
of the formation
enthalpies for monovacancies and for charge-impurity point-defects  
\cite{jo99,jo01} produced values which showed that the assumption
could not be valid.  In that work, a simple procedure was used to
obtain the properties of homogeneous lattices as a basis from which
the enthalpy changes produced by perturbations (the point-defects)
could be calculated.  Nuclei were described by the
compressible liquid-drop model (CLDM) of Lattimer {\it et al}
\cite{lpr85} with their model parameter values.  Table I contains
the calculated lattice parameters used.  The CLDM nuclear charge
$\bar{Z}$ is
a continuous variable, and the neutron and proton chemical potentials
are given with reference to the neutron rest energy.  The formation
enthalpy for a point-defect of charge $Z$ can be expressed as
$H_{FZ}=C(Z -\bar{Z})^{2}$ relative to the homogeneous lattice.  It is
reasonable to expect that the order of magnitude of $H_{FZ}$
should be related with the monovacancy formation enthalpy $H_{Fv}$ 
through the expression $H_{FZ} \approx (Z\bar{Z}^{-1} - 1)^{2}H_{Fv}$,
with $C \approx H_{Fv}\bar{Z}^{-2}$.
The values of $C$ contained in the final column are broadly consistent
with this, given that $H_{Fv}\approx 15$ MeV \cite{jo99}.  These
$H_{FZ}$ values, determined solely by Coulomb lattice and bulk nuclear
matter properties, were small and it became obvious that shell
corrections should not be neglected.  Their inclusion
\cite{jo04}, based on the single-particle levels found by Negele and
Vautherin \cite{nv73} and the Strutinski procedure (see Ref. \cite{rs80}),
changes the $H_{FZ}$ to the formation enthalpies shown in Fig. 1.  These
are exclusive of integral multiples of the electron and neutron chemical
potentials.  The quadratic dependence
on $Z - \bar{Z}$ and the effect of proton closed shells are both obvious.
We refer to Ref. \cite{jo04} for further details of the methods of
calculation of these $H_{FZ}$ and of the approximations and assumptions on
which they are based.  They are necessarily subject to some uncertainty
and it is also the case that, for laboratory nuclei far from stability,
the dependence of the
spin-orbit interaction on neutron excess remains a problem of current
interest (see Schiffer {\it et al} \cite{sch04}).  Although the formation
enthalpies given in Fig. 1 are, as we have stressed, of uncertain
quantitative value (which is why the simplification of approximating
free energy by enthalpy has been made), it is
fairly certain that shell corrections producing local minima in $H_{FZ}$
must exist, even if they are smaller than those of \cite{jo04}.
An additional factor supporting this conclusion is the belief  
that the quadratic dependence of $H_{FZ}$ on $Z - \bar{Z}$ has
not been seriously under-estimated.  (The reason for this is that the
monovacancy formation enthalpies obtained in \cite{jo99} are large when
expressed in units of the melting temperature,
$H_{Fv} \approx 35 k_{B}T_{m}$, whereas for alkali metals they are
$\approx 10 -15 k_{B}T_{m}$.) 
Therefore, the qualitative features of Fig. 1, and specifically the
existence of large potential barriers between closed-shell values of
$Z$, are probably well-founded.
The consequence is that weak-interaction equilibrium fails during
cooling of primordial neutron-star matter \cite{jo04} producing a
$Z-$heterogeneous and amorphous solid.  This is the primordial state
of the solid in an isolated neutron star or in a
binary system that has not yet evolved to a high rate of mass accretion.

\begin{figure}
\includegraphics[width=84mm]{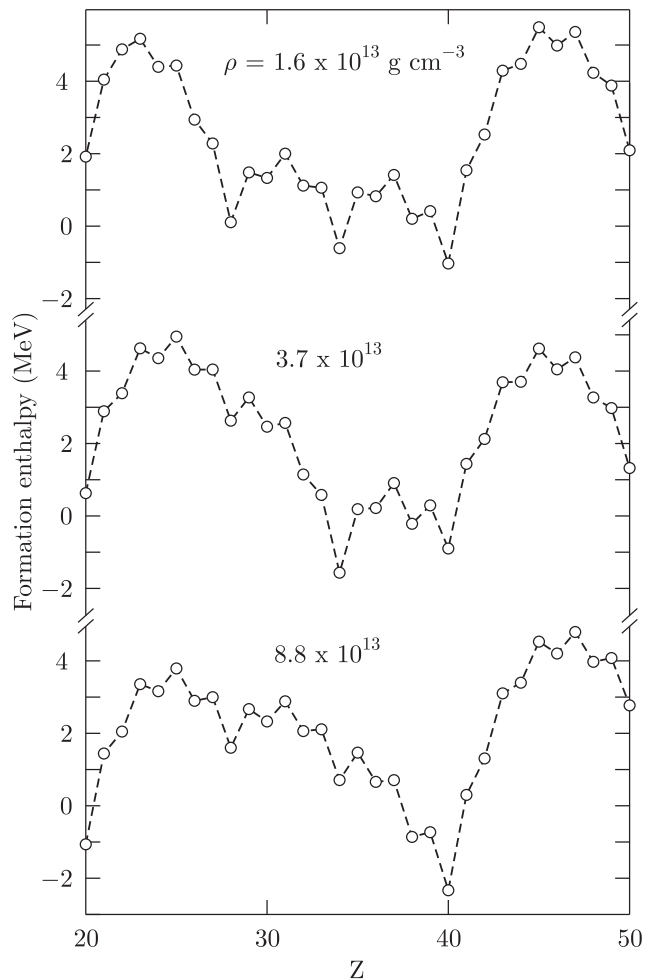}
\caption{\label{Fig. 1} Formation enthalpies for charge-$Z$ point defects,
from the work of \cite{jo04}, are shown for the homogeneous lattices
whose properties are given in Table I.  Integral multiples of the electron
and neutron chemical potentials are excluded.  There are local minima at
the proton closed-shells $Z= 20, 28, 34, 40, 50$.}
\end{figure}
 
\begin{table*}
\caption{\label{I}Homogeneous CLDM lattice properties are given for a set
of matter densities $\rho$.  These are the neutron continuum number
density $n^{e}_{n}$, the nuclear charge $\bar{Z}$ and radius $r_{N}$, the
Wigner-Seitz radius $r_{WS}$, the neutron and proton chemical potentials
$\mu_{n,p}$ and the pressure $P$.  Also given is the CLDM proton chemical
potential $\mu^{e}_{p}$ defined in the limit of zero proton number
density within the neutron continuum.  The final two columns give the
proton barrier penetration factor $\exp{(-\kappa x)}$ and the parameter
$C$.}
\begin{ruledtabular}
\begin{tabular}{c c c c c c c c c c c} \\
$\rho$ & $n^{e}_{n}$ & $\bar{Z}$ & $r_{N}$ & $r_{WS}$ & $\mu_{n}$ &
$\mu_{p}$ &
 $\mu^{e}_{p}$ & $P$ & $\exp{(-\kappa x)}$ & $C$ \\
 ($10^{13}$ g cm$^{-3}$) & ($10^{-3}$ fm$^{-3}$) &   & (fm) & (fm) & 
 (MeV) & (MeV) & (MeV) & (MeV fm$^{-3}$) &   & (MeV) \\
\hline \\
 1.6 & 7.8 & 34.6 & 5.8 & 27.1 & 4.9 & -46.4 & -9.1 & 0.0176 & 
 2.0$\times 10^{-25}$ & 0.0142  \\
 3.7 & 18.4 & 35.1 & 6.3 & 22.0 & 7.2 & -53.6 & -20.9 & 0.0526 &
  7.2$\times 10^{-18}$ & 0.0096  \\
 8.8 & 43.6 & 34.3 & 7.1 & 17.1 & 10.1 & -63.7 & -45.7 & 0.158 &
  7.2$\times 10^{-9}$ & 0.0051  \\
\end{tabular}
\end{ruledtabular}
\end{table*}

\subsection{\label{B}Elimination of heterogeneity by 
quantum-mechanical tunnelling}

The metastable state of matter described in Sec. IIA is not possible at
matter densities approaching the phase transition to the core liquid.
Lower-dimensional phases, if they exist at densities between those of the
spherical nuclear phase and liquid core \cite{pr95}, are unable to
support point-defects.  But since such phases may not exist, it is
necessary to investigate the stability of the $Z-$heterogeneous solid
for values of $\rho$ larger than those of Table I.  At finite
temperatures, quantum-mechanical tunnelling of single protons and of
pairs can occur between any nucleus and its nearest neighbours.
Consider two nuclei with proton closed shells $Z_{1,2}$.  The
tunnelling of a single proton forms particle and hole excitations,
$Z_{1}\rightarrow Z_{1}\pm 1$ and
$Z_{2}\rightarrow Z_{2}\mp 1$ with an associated enthalpy
change $\delta H_{F} > 0$ which must be balanced by interaction with
other degrees of freedom present in the system.  (Pair
tunnelling would allow formation of even$-Z$ nuclei but with much smaller
transition rates.)  The nature
of the potential barrier in this unusual system is somewhat uncertain,
but we have assumed that it is determined by the difference between
that proton chemical potential $\mu^{e}_{p}$ which is defined by
proceeding to the limit of
zero proton number density within the neutron continuum, and the
chemical potential $\mu_{p}$ in the nuclear volume.  The values of
$\mu^{e}_{p}$ given in Table I have been obtained from the Skyrme
bulk nuclear matter pseudo-potential used by Lattimer {\it et al} 
\cite{lpr85}.  The order of magnitude of the barrier penetration
factor, calculated for the homogeneous lattice of charge
$\bar{Z}$, in which
$\hbar^{2}\kappa^{2} = 2m_{p}(\mu^{e}_{p}-\mu_{p})$ and
$x=2(r_{WS}-r_{N})$, is given in Table I.

Scattering of neutron superfluid quasiparticles by protons
is the most important interaction with other degrees of freedom.
A very approximate estimate of the
transition rate for the tunnelling process $Z_{1}\rightarrow Z_{1}\pm 1$
and $Z_{2}\rightarrow Z_{2}\mp 1$ can be found from the Table I barrier
penetration factor.  It is of the order of
\begin{eqnarray}
\gamma_{q} = \frac{m_{n}^{2}k_{Fn}^{2}k_{B}T \left|\bar{V} \right|^{2}}
{2 \pi^{3}\hbar^{5}}
{\rm e}^{-2\kappa x-\beta\delta H_{F}-\beta\Delta_{n}}
\end{eqnarray}
where $\beta^{-1}=k_{B}T$ and the parameters $k_{Fn}$ and $\Delta_{n}$
are the neutron Fermi wavenumber and energy
gap within the nuclear volume.  The order of magnitude of the
strong-interaction
scattering matrix element is $\bar{V} \sim 10^{2}$ MeV fm$^{3}$, leading to a
transition rate
\begin{eqnarray}
\gamma_{q} \sim 10^{19}
{\rm e}^{-2\kappa x-\beta\delta H_{F}-\beta\Delta_{n}} {\rm s}^{-1},
\nonumber
\end{eqnarray}
at $10^{9}$ K.
For temperatures expected either during accretion or early in the
life of an isolated neutron star, and for values $\delta H_{F} \sim 1$ MeV
that are
consistent with Fig. 1, the most significant exponential term in Eq. (1)
is certainly the barrier penetration
factor. Its rapid increase from negligibly small values is an unavoidable
consequence of the variations of both
$r_{WS}$ and $\mu^{e}_{p}-\mu_{p}$ with increasing $\rho$.
At values of $\rho$
not much larger than the third row of Table I, the proton tunnelling
transition rate must become so large that the metastable equilibria
represented by the minima in Fig. 1 become short-lived and
all nuclear charges approach that of the most favoured closed
shell, probably $Z=40$ \cite{nv73}.

Our conclusion is that the state of matter at this density and beyond
is independent of its past history and that, in consequence, any direct
transition from the spherical nuclear phase to the
liquid must occur at a unique pressure $P_{c}$ and density $\rho_{c}$.
The same statement can be made of any intermediate transition to
lower-dimensional solid phases which can support some anisotropic
stress-tensor components \cite{pp98}. This has the consequence that,
within the Cowling approximation,
in which the perturbation to the gravitational potential caused by an
angle-dependent solid crust composition is neglected, the static
structure of the neutron star is very simple.  The surface separating
liquid and solid phases is necessarily
isobaric and spherical with radius $R_{c}$,  and these conditions are
maintained by strong and electromagnetic interactions. This statement,
of course, neglects rotation and the effect of thermal fluctuations which
average to zero over macroscopic areas.  The extent to which it would
be true for the various modes of oscillation of a neutron star is
not considered here.

\section{\label{III}Evolution of accreted matter}
\subsection{\label{A}The inward flow of matter}

The predicted nuclear mass numbers produced in the {\it rp} process
burning of accreted matter extend to $A \approx 110$
\cite{bb98, sbcw99, sabb01, khkf04}.  Widths and shapes of these
mass number distributions
depend on thermodynamic conditions which are determined principally by
the local rate of accretion and its hydrogen-helium composition.
Therefore, it is very likely that the mean mass number of nuclei entering
the surface ocean of the star is a function of surface position specified
by the angles $\theta,\phi$, with the possibility
of significant differences between the magnetic polar and equatorial
regions.  The extent to which convective motions exist in the ocean
is a complex problem but is known to
depend on magnetic field strength and on the presence of a non-radial
temperature gradient \cite{muv97, vu04}.  The more conservative
assumption is that, in general, mixing is
incomplete over the whole surface area and that some composition
angle-dependence remains as matter enters the solid phase of the crust.
This inward movement of individual nuclei occurs at constant $A$, but
with $Z$ decreasing owing to electron capture transitions.
Given the width of the
predicted distributions, the solid structure formed must be amorphous,
with any order limited to very short lengths \cite{jo04}.
  
However, further structural rearrangement must occur at a very high
rate during compression because the average nuclear cell radius
decreases during the inward movement.
Below the neutron-drip threshold, it is $\propto \rho^{-1/3}$ and at
$\rho > \rho_{nd}$, the density-variation obtained from column 5
of Table I shows almost the same dependence.
These variations are very much more rapid than those which accompany
the inward radial movement of a lattice without change of form and
necessarily imply that structural rearrangement occurs.
Its effect can be seen by assuming that, at
some instant of time, the strain components are $\epsilon_{ij}$,
defined relative to suitable fixed coordinates.
Given the $\rho$-dependence of $r_{WS}$, it
follows that nuclear rearrangments consequential to inward movement
can, in principle, change the strain at a rate as large as
$\dot{\epsilon}_{ij}\sim \tilde{v}/L_{\rho}$ where
$L_{\rho}\sim 10^{4}$ cm is the
scale-length for radial density variation and $\tilde{v}$ is the
bulk inward radial velocity of matter at density $\rho$. Although
we are unaware
of any strictly analogous terrestrial system, the high rate of
nuclear rearrangement has some similarity with visco-elasticity.
The solid system considered here must have the same
elastic stress-response as an amorphous solid for times several
orders of magnitude shorter
than $\epsilon_{ij}L_{\rho}/\tilde{v}$, but
over longer times, the strain is continuously
changing owing to the inward movement of accreted matter. 
The important feature is that the solid, unlike
a conventional visco-elastic structure, has no intrinsic
relaxation time: the values of $\dot{\epsilon}_{ij}$ are determined
by the accretion rate.  Later in the evolution of the binary
system, when the accretion rate has become neglible, the
structure is again that of an amorphous solid.

Below $\rho_{nd}$, the equation of state of neutron star matter is
polytropic, $P = K\rho^{\Gamma}$,  with an adiabatic index $\Gamma$
very close to $4/3$ because the pressure $P$ is well approximated
by the relativistic electron pressure $P_{e}$.  In this region, an
incremental change in nuclear composition from reference mean values
$\bar{A}$ and $\bar{Z}$ produces a simple change in
the constant $K$.  At a fixed $P_{e}$, the matter density
$\rho \propto \bar{A}/\bar{Z}$ to a high level of approximation.
The adiabatic index is constant and therefore,
\begin{eqnarray}
\frac{\delta K}{K} = - \Gamma \frac{\delta(\bar{A}/\bar{Z})}
{(\bar{A}/\bar{Z})}.
\end{eqnarray}
No similar elementary result can be derived
at $\rho > \rho_{nd}$ because the neutron continuum contributes the
greater part of the pressure.  But the CLDM equation of state underlying
Table I is approximately polytropic, with adiabatic index
$\Gamma = 1.29$, and we shall assume that composition changes in this
region also give $K \rightarrow K + \delta K$ with no incremental
change in $\Gamma$.

\subsection{\label{B}Angle-dependent structure of the solid crust}

Given the considerations of Sec. IIB and IIIA, the basic properties
of the crust in an accreting neutron star can be found very simply using
a polytropic equation of state and the Cowling approximation, in which
perturbations of the gravitational potential are neglected.
Suppose that the nuclear composition is angle-dependent, giving
an equation of state with $K \rightarrow K + \delta K(\theta,\phi)$,
but with the length-scale for non-radial variation of $\delta K$
at least of the same order of magnitude as the crust depth.
A three-dimensional elastic-deformation treatment of the crust is
appropriate for calculations of the adiabatic modes of oscillation
such as those of McDermott, Van Horn and Hansen \cite{mhh88} at zero
or negligible accretion rate.  Nevertheless, given the many
uncertainties in the formulation of the present problem,
a one-dimensional (radial) approximation for local values of
$\theta, \phi$ has the merit of transparency and is adequate.

In the three-dimensional treatment, Eulerian deviations from the
reference system ($\delta K =0$) satisfy the full static equilibrium
condition (or quasi-static in view of the accretion flow)
\begin{eqnarray}
\frac{\partial \delta\sigma_{ij}}{\partial x_{j}}+
\delta\rho g_{i} = 0,
\end{eqnarray}
in which the deviation in the stress tensor,
\begin{eqnarray}
\delta\sigma_{ij} = \delta\sigma^{sh}_{ij}+\sigma^{M}_{ij}-
\delta P\delta_{ij},
\end{eqnarray}
is the sum of the deviation in the shear components of the elastic
stress tensor, the Maxwell tensor, and the term
$-\delta P$ which includes the deviations in all other isotropic
stress tensor components of the system. Let us
assume that $\delta K \neq 0$ within a pressure interval
$0 < P < P_{a}$, where $P_{a}\leq P_{c}$. 
(It is possible that $P_{a}$ may be slightly composition and
therefore angle-dependent,
but this would have an effect one or more orders of magnitude smaller
than those of first order in $\delta K$.)
In the interval $P_{a} < P < P_{c}$ where matter has reached
equilibrium, either through high rates of proton tunnelling or a
transition to a lower-dimensional structure, the angle-dependence
vanishes so that $\delta K = 0$. 
Eq. (3) satisfy the boundary conditions $\delta\sigma_{rr} = 0$
at pressures $P= 0, P_{a}$, though possibly with a small error if
$P_{a} < P_{c}$.  The density deviation, integrated from
$R_{a}$ to the surface of the star is then
\begin{eqnarray}
\int^{R}_{R_{a}}\delta\rho dr = -\frac{1}{g_{r}}\sum_{j\neq r}
\int^{R}_{R_{a}}\frac{\partial\delta\sigma_{rj}}{\partial x_{j}}dr.
\end{eqnarray}
In a true one-dimensional approximation, the integrated density
deviation therefore vanishes.  It is worth comparing Eq. (5) with
Eq. (51) and (52) of Ref. \cite{ucb00} which satisfy identical
boundary conditions.  But these authors find, by reference to their
three-dimensional solutions, that
$\mid\delta\sigma_{rr}\mid$ is at least an order of magnitude 
larger than $\mid\delta\sigma_{rj}\mid$
throughout most of the interval of integration.  They therefore
conclude
that a one-dimensional estimate of a quadrupole moment which they
express, in Eq. (52) of their paper, as
\begin{eqnarray}
Q_{22} = R^{4}\int\delta\rho dr
\end{eqnarray}
is extremely poor owing to its independence from the major stress
deviation $\delta\sigma_{rr}$. It is obvious that this is correct.
(In the one-dimensional approximation which will be adopted here, the
moment
so defined would be $Q_{22} = 0$.) These authors observe that changes in
the radial distribution of density produced by $\delta K \neq 0$ are a
far larger contributor to quadrupole moments than the non-radial
displacements which also arise.
Although the latter can be large ($\sim 10^{4}$ cm), they are smaller
than the typical scale length for non-radial variation of $\delta K$ and
so are much less significant.
For a thin crust, it must be the
case that any calculation which gives the correct radial distribution
of density deviations also gives a satisfactory approximation to
the quadrupole moment. A one-dimensional solution of the form given
here in Eq. (7) and (8) is capable of this.  But it is necessary to use
the correct quadrupole moment definition given by Eq. (1) of
Ref. \cite{ucb00} instead of Eq. (6).

Solutions of Eq. (3) are easily obtained in the one-dimensional
approximation for the case in which $\delta K$ is
independent of depth within the pressure interval $0 < P < P_{a}$.
We replace $\delta\sigma_{rr}$ by $-\delta P$ and find that
the incremental changes from the
reference values of pressure and density are,
\begin{eqnarray}
\delta P^{(1)} & = & \frac{P\delta K}{K(\Gamma -1)}\left(\left(
\frac{\rho_{a}}
{\rho}\right)^{\Gamma -1} - 1 \right), \nonumber \\
\delta \rho^{(1)} & = & \frac{\rho \delta K}{K\Gamma (\Gamma -1)}
\left(\left(\frac{\rho_{a}}{\rho}\right)^{\Gamma -1} - \Gamma \right).
\end{eqnarray}
The radial displacement of a point fixed in Lagrangian coordinates is,
\begin{eqnarray}
\delta r^{(1)} = \frac{\delta K}{g_{a}(\Gamma - 1)}\left(\rho_{a}
^{\Gamma - 1} - \rho^{\Gamma - 1}\right),
\end{eqnarray}
where $g_{a}$ is the magnitude of the gravitational acceleration
${\bf g}$ at pressure $P_{a}$.  There is a density discontinuity
$-\rho_{a}\delta K/K\Gamma$ at $P_{a}$.  The radius of the
stellar surface $(P = 0)$ is then angle-dependent and the case
$\delta K > 0$, a slightly more stiff equation of state, produces
a local increase in radius.
 
Eq. (4) makes
obvious the fact that the term $\delta P$ produced by an angle-dependent
$\delta K \neq 0$ has an effect analogous with the isotropic components
of the Maxwell tensor.   
The change of phase to the liquid occurs at a fixed pressure
$P_{c}$ and density $\rho_{c}$.  In the Cowling approximation, and
neglecting the effect of rotation, the
radius $R_{c}$ is angle-independent, but its magnitude reduces very
slowly with time for most equations of state during accretion
as the total mass of the
star increases.  Solution of the three-dimensional Eq. (3) has not been
attempted.  It would not be simply a problem in elasticity because
$\epsilon_{ij}$ is time-dependent for non-zero accretion rates as we
have described in Sec. IIIA.  From Eq. (3) and (4), and with neglect of
the Maxwell tensor, we can see that the non-radial components of the
shear stress deviation $\delta\sigma^{sh}_{ij}$ are of the same
order of magnitude as $\delta P$.  Hence the maximum $\delta P$ for
which the solid will respond quasi-elastically is of the order of
$10^{-2}\tilde{\mu}$, for an amorphous structure, where $\tilde{\mu}$
is here the shear modulus.  For greater values of $\delta P$, the
behaviour of the system will approach that of a fluid.  The very
complex problem represented by this condition is not considered here.
In the neutron-drip region, the pressure and shear modulus have almost
identical $\rho$-dependences and so are approximately related by
$\tilde{\mu}\approx 6\times 10^{-3}P$.  Therefore, the maximum
$\delta K$ for which the solid behaves quasi-elastically is given
by $\delta K/K \approx 10^{-2}\tilde{\mu}/P \approx 6\times 10^{-5}$.
We can see that the inward movement of matter during accretion is basically
radial but, given the non-radial gradient in $\delta P$, it cannot be
exactly so.

\subsection{\label{C}Nuclear transitions above the neutron-drip threshold}

The neutron-drip region at $\rho > \rho_{nd}$ contains almost all the
mass of the crust.  Some information about the distribution
of $A$ and $Z$ for matter moving into this region has been provided by
Haensel and Zdunik (see Fig. 1 of \cite{hz03}).  Successive electron
captures in the outer crust, for the two mass number examples $A=56, 106$
considered by these authors, lead to the formation of nuclei with
charges $Z=18,32$, respectively, as the inward moving accreted matter
is compressed to $\rho_{nd}$.  The assumption which
appears to have been made by Sato \cite{sat79} and by later authors
\cite{hz03}
is that, with further compression inside the neutron-drip region,
electron capture transitions
continue at constant $A$ until the transition rates for pycnonuclear
fusion reactions become appreciable at $Z\approx 10$.  Comparison with
calculations of the neutron-drip state, for example, those of Negele
and Vautherin \cite{nv73}, shows that the charges found by Haensel
and Zdunik ($Z=18,32$) lie well below the equilibrium $Z$.  Therefore,
unless some special case for the contrary can be made, the nuclear
transitions which occur must follow a path of increasing $Z$.  They are,
principally,
the successive capture of neutron pairs followed by electron emission.
For $\mu_{n} > 0$, nuclei can be viewed as bound states of protons
embedded in a neutron continuum.  Thus there is no barrier to
increasing nuclear $A$ and $Z$ values by these transitions.
Pycnonuclear fusion rates are negligible except possibly at the
low-$Z$ end of the nuclear number density distribution.

The work of Ref. \cite{hz03} shows that the inward moving accreted
matter may have a fairly wide distribution of $Z$ at $\rho_{nd}$,
but with an rms value $Z_{a}$ which is probably much smaller than either
the equilibrium values calculated by Negele and Vautherin \cite{nv73}
or those given in Table I.  In order to see how the accreted matter
changes as it is further compressed to $\rho > \rho_{nd}$ it is
convenient to use, with some modifications, the CLDM procedure
which is the basis for Table I.  Strong and electromagnetic-equilibrium
constraints are retained, but the
condition $\delta\mu_{i} = 0$, where
\begin{eqnarray}
\delta\mu_{i} = \mu_{n}-\mu_{pi}-\mu_{e}-(m_{p}-m_{n})c^{2}
-\frac{\partial f_{c}}{\partial Z}
\end{eqnarray}
and $f_{c}$ is the Coulomb energy of the Wigner-Seitz cell \cite{lpr85},
is removed so that the nuclear charge $Z$ can be treated as a
constant parameter.  For the modified CLDM procedure, the
distribution of $Z$-values is approximated
by a binary system $(i=1,2)$.  The proper-frame neutron chemical
potential $\mu_{n}$ is related with the time-like component of the
metric and so, to a good approximation, can be assumed a
fixed quantity at any point.  However, the electron chemical
potential $\mu_{e}$, defined here as the Fermi energy,
necessarily has a common value for the binary system which has
to be determined locally.  This has been achieved by modifying
the constraint \cite{lpr85} relating the nuclear surface
thermodynamic potential density, defined here as $\tilde{\sigma}_{i}$,
with the Coulomb
energy of the Wigner-Seitz cell.  (In the binary system and as a matter
of convenience, we refer
to the electrically neutral sphere, with nucleus $Z_{i}$ at the
origin, as the Wigner-Seitz cell.)  Clearly this cannot be satisfied
simultaneously for the binary mixture $i=1,2$ and so we have adopted
the form
\begin{eqnarray}
\sum_{i}a_{i}\left(2\pi r_{Ni}^{2}\tilde{\sigma}_{i} - f_{ci}\right) = 0,
\end{eqnarray}
for this constraint, where $r_{Ni}$ is the nuclear radius and $a_{i}$
the number density fraction .  These are the two modifications to the
CLDM procedure that have been made.  The two remaining constraints
in the CLDM procedure, those of neutron chemical potential and 
nuclear pressure equilibria are satisfied by both nuclei of the binary
system.

Although the modification given by Eq. (10) is a very elementary
approximation for the binary system, the most important strong and
electromagnetic-interaction constraints have been retained.
Calculations of the equilibrium, using this CLDM procedure,
confirm that the basic properties of
accreted matter change in an intuitively obvious way as it is
compressed to $\rho > \rho_{nd}$. For
the case in which the binary system mean square nuclear charge
$\sum a_{i}Z^{2}_{i} = \bar{Z}^{2}$, where $\bar{Z}$ is the
equilibrium charge given in Table I (for the lowest matter density, and
with $Z_{1}=28$ and
$Z_{2}=40$), we find that the computed values of
$\delta\mu_{i}$ are very small and that there are
negligible changes in $P$, $\rho$ and chemical potential per
baryon.  This is consistent with the discussion
of formation enthalpy differences given in Sec. IIA.  But the mean
square nuclear charge $Z_{a}^{2}$ of accreted matter at
$\rho_{nd}$ is smaller \cite{hz90,hz03} than $\bar{Z}^{2}$.
Therefore, we have computed binary model cases in which $Z_{1}=\bar{Z}$
and $Z_{2}=20$ and find that the chemical potential imbalance defined
by Eq. (9) is an approximately linear function of $a_{2}$, and is
given by $\delta\mu_{2} \approx 0.6 + 27a_{2}$ Mev.
The implication is that weak-interaction transitions rapidly
reduce the value of $a_{2}$ at $\rho > \rho_{nd}$ and change the
system toward the equilibrium state of Table I.  Ultimately, for small
values of $\bar{Z} - Z_{a}$ and therefore of $a_{2}$, the nuclei
of charge $Z_{2}$ can be regarded as impurities in the $\bar{Z}$
system so that the
formation enthalpy distribution of Fig. 1 becomes valid and the
speed and efficiency of this process are determined by the
temperature and by the formation enthalpy
barriers between the closed shells at $Z=20,28,34$ described
in Sec. IIA.

Nuclear reactions caused by compression of the accreted matter
are a volume source of heat. Ref. \cite{ucb00} gives temperature
distributions for an accretion rate of $10^{-8}M_{\odot}$ yr$^{-1}$
calculated using various assumptions about neutron superfluidity.
They are fairly slowly varying functions of $\rho$ with maxima of
$7-8\times 10^{8}$ K at $\rho_{nd}$.  (These high values show
that a classical neutron gas must be present in a finite region with
$\mu_{n} < m_{n}c^{2}$.)  For this accretion rate, the inward radial
velocity is $6\times 10^{-8}$ cm s$^{-1}$, at the threshold
$\rho_{nd} = 6\times 10^{11}$ g cm$^{-3}$ \cite{hz90}.
Initially, for large
imbalances $\delta\mu_{i} > 0$, weak-interaction transitions have
no formation enthalpy barrier and are rapid.  But as the value of $Z_{a}$
increases toward  $\bar{Z}$ and the formation enthalpy distribution of
Fig. 1 becomes valid, the question
arises of the extent to which thermal excitation of the weak
transitions allows movement between proton closed-shells. 
Low enough formation enthalpy barriers would allow all nuclei
to reach a unique proton closed-shell well within the transit
time of accreted matter in the crust.

This is an important question because the modified CLDM procedure
described above shows that a nuclear charge imbalance,
of the form $Z_{a} < \bar{Z}$, has a significant effect on the equation
of state.  This should not be unexpected because both the
nuclear pressure equilibrium constraint and Eq. (10) depend
quadratically on $Z$.
For the binary system charges $Z_{1}=\bar{Z}$ and
$Z_{2}=20$, at the lowest density of Table I, the incremental changes
$\delta P$ and $\delta \rho$ with respect
to the system with mean charge $\bar{Z}$ give an incremental change
in the equation of state,
\begin{eqnarray}
\frac{\delta K}{K} = \frac{\delta P}{P} - \Gamma\frac{\delta\rho}{\rho}
= 0.15a_{2},
\end{eqnarray}
for an assumed constant $\Gamma= 4/3$.  In terms of quadrupole-moment
generation, this
would be a potentially enormous effect if its presence were
angle-dependent.  The problem becomes one of estimating
the interval of $\rho$ over which $a_{2}$ is non-negligible.

Assuming the formation enthalpies given in Fig. 1 (lowest density),
the rates for $Z=23\rightarrow 24$ and $Z=25\rightarrow 26$ have
been calculated as functions of temperature.  The approximations made in
calculating weak-interaction transition rates are precisely those
described in previous papers \cite{jo01,jo04}.  In the temperature
region of interest, the time-constant for the depletion of a small
value of $a_{2}$, with $Z_{2}=20$, is $\tau \approx 3.3\exp(51/T_{9})$ s,
where $T_{9}$ is the temperature in units of $10^{9}$ K.
The times, for the temperatures predicted at $\rho_{nd}$ in
Ref. \cite{ucb00},
are so long that there is no question that significant non-zero
$a_{2}$ can survive the $\sim 10^{14}$ s interval of rapid accretion.
But we have previously stressed the many sources of uncertainty in
the details of Fig. 1.  The existence of these problems is unfortunate
because halving the formation enthalpy barrier (roughly equivalent to
halving the exponent in the time-constant expression) reduces the
depletion time constant to values within one or two orders of
magnitude of the accretion time-interval. 

In summary, the evolution of accreted matter in the neutron-drip
region is as follows.  A nucleus of charge $Z$ at $\rho_{nd}$
undergoes rapid weak-interaction transitions, with capture or
emission of neutron pairs, toward the most accessible proton closed
shell configuration.  The calculations of burning
\cite{bb98, sbcw99, sabb01, khkf04} and of evolution below the
neutron-drip threshold \cite{hz03} indicate
$Z$-distributions which would evolve by transitions
toward those at $Z=20, 28, 34$ and possibly $40$.  Their relative
populations reflect the width and form of the distribution of $A$
formed in the atmosphere.  But the mean square charge $Z_{a}^{2}$
at $\rho_{nd}$ is appreciably smaller than $\bar{Z}^{2}$ so that,
as our CLDM binary system calculations confirm, there will be
further rapid transitions toward one or more closed shells near
$\bar{Z}$ until the populations of the other closed shells become
small, of the order of $a_{i}\sim 10^{-2}$.  At this stage, the
system may be metastable or, depending on temperature and on
formation enthalpy barrier height, there may be further transitions
to a unique closed-shell $Z$.  As we have emphasized in the previous
paragraph, it is unfortunately the case that we cannot decide between
these two possibilities with any confidence.  The sequence of
formation enthalpy distributions shown in Fig. 1 indicate that
compression to $\rho > \rho_{nd}$ may itself change the closed-shell
populations.  Thus the $Z= 28$ minimum
becomes less pronounced in the vicinity of $3.7 \times 10^{13}$
g cm$^{-3}$ and it
is possible that its population is transferred firstly to $Z= 34$
and then, near $8.8 \times 10^{13}$ g cm$^{-3}$, to $Z= 40$.  But it
must be re-emphasized that the specific values of $H_{FZ}$ in Fig. 1
should not be taken too seriously \cite{jo04} and that although
they show the general way in which the system evolves, any details
that have been presented for illustrative purposes are not 
necessarily reliable.  These changes under compression continue
until the density reaches $\rho_{a}$, where the transition rates for
quantum-mechanical proton tunnelling become so large that any
metastability disappears.  

Non-radial temperature gradients, if present in the solid crust, are an
independent source of composition angle-dependency on any isobaric
surface \cite{bil98,ucb00} at $\rho< \rho_{nd}$.  This arises, in
principle, because the weak-interaction transition rates which change
$Z$ as $\rho$ increases are temperature-dependent. Above $\rho_{nd}$,
the initial progression of a nucleus toward a particular proton closed
shell is rapid because there is no formation enthalpy barrier. Thus
temperature is unlikely to be a significant factor.  But sequences of
transitions between adjacent closed shells are, as we have described,
extremely temperature-dependent and it is not possible
to predict these rates with any confidence.

\section{\label{IV}Conclusions}

Unfortunately, the main conclusion reached in this paper is that it
is not possible to calculate the angle-dependence of the equation
of state (the incremental function $\delta K(\theta,\phi)$) arising from
either an angle-dependent atmospheric composition or non-radial
temperature gradients in an accreting neutron
star.  We have shown that such a calculation depends on nuclear
structure properties that are very unlikely to be established with
any degree of certainty in the immediate future.  But this conclusion,
though unfortunately negative, is not without value in relation to
future developments in gravitational wave detection.

The hypothesis that the limiting rotation frequencies of neutron stars
in low-mass X-ray binary (LMXB) systems are determined by gravitational
radiation \cite{bil98, ucb00} specifies a fairly compact range of
values for the mass-quadrupole tensor component.
Integration of either Eq. (5) or (7) shows that
an increment $\delta K \neq 0$ produces no change in the total
mass within the interval $0 < P < P_{a}$ though there is a density
discontinuity $\delta\rho_{a}$ at $P_{a}$.  Therefore,
the resultant mass-quadrupole tensor component, as defined by Eq. (1)
of Ref. \cite{ucb00} with $\delta\rho_{lm}$ replaced by
$\delta\rho^{(1)}$, arises solely from the mass rearrangement 
given by Eq. (7) and is
\begin{eqnarray}
Q_{22} \approx \frac{3}{20}h^{2}\rho_{a}R_{a}^{3}\frac{\delta K}{K},
\end{eqnarray}
where $h$ is the depth of the $\delta K \neq 0$ layers.  If these are
confined to $\rho < \rho_{nd}$, for layers such that
$h^{2}\rho_{a} \approx 10^{19}$ g cm$^{-1}$, the predicted
mass-quadrupole tensor is several orders of magnitude smaller than the
specified value, $Q_{22} = 3.5\times 10^{37}$
g cm$^{2}$ \cite{ucb00}.  In the neutron-drip region, values of
$h^{2}\rho_{a} \approx 10^{24}$ g cm$^{-1}$ are possible, for which the
specified $Q_{22}$ would be given by
$\delta K \approx 2\times 10^{-4}K$.  It is also interesting that these
$\delta K$ are not much different from the approximate value corresponding
with the quasi-elastic limit noted at the end of Sec. IIIB.  This is
not inconsistent with the conclusions of Ref. \cite{ucb00}.  With
reference to Eq. (11), we can see that only a very small charge
imbalance, if suitably angle-dependent, is needed to produce these
$\delta K$ at a given density.
But, as we have emphasized, it is not possible to predict whether
or not the metastable population of proton closed shells can be
sufficiently long-lived to maintain this.

The above conclusions are for angle-dependent composition asymmetries
formed in the atmosphere of the star and make no reference to
composition asymmetries produced by non-radial internal
temperature gradients \cite{bil98,ucb00} which may be present.
Those formed at $\rho < \rho_{nd}$ have too small an effect to
be of significance \cite{ucb00}, but at $\rho > \rho_{nd}$, there
is a possibility that the conclusions of Sec. IIIC may require
modification.  The value of $Z_{a}$ at $\rho_{nd}$ may have a small
temperature-dependence, but this will not affect the population
of proton closed shells initially formed.  The major effect of
temperature asymmetry is in changing the transition rates from
these closed shells to the closed-shell $Z$ of complete weak-interaction
equilibrium.  As emphasized
in Sec. IIIC, these rates are exponentially-dependent on formation
enthalpy differences that are not well-known.  Although they are
probably many orders of magnitude too small or large for conceivable
temperature gradients to have any effect, there is a  
non-quantifiable though small probability that transition rate
orders of magnitude may allow the formation of angle-dependent
proton closed-shell populations within an appreciable interval
of $\rho$.

It is worth considering briefly the extent to which intrinsic
temperature-dependence of the equation of state can lead to
asymmetry in the presence of non-radial temperature gradients.
There appears to be no published work on this problem for the
case of the solid phase of neutron star matter.  Calculations
at high temperatures \cite{lpr85} assume a normal neutron continuum,
also the presence of classical (translational) degrees of freedom,
and so cannot be extrapolated to $T < T_{m}$.  In this region, the
most important source is likely to be the temperature dependence of
the electron partial pressure $P_{e}$.  For the order of
magnitude of temperature variation assumed in Ref. \cite{ucb00}
($\delta T\sim 0.05T$ at $T\approx 4\times 10^{8}$ K), the resultant
increment of $\delta K \sim 4 \times 10^{-8}K$
at a typical chemical potential $\mu_{e} = 60$ MeV is too small
to be of significance.

The origin of the differences between this paper and Ref. \cite{ucb00}
have been described in Sec. III.  The evolutionary path of accreted
matter at $\rho > \rho_{nd}$ is the major area of disagreement.  Our
analysis is that it is determined by the proton closed-shell structure
of nuclei.  The disagreement concerning the treatment of the solid
crust is much less important and in Sec. IIIB we have shown how it arises.

A three-dimensional calculation of the structure for an angle-dependent
equation of state forms a large part of Ref. \cite{ucb00}.
These authors assume an elastic response and the analysis given draws
on that of McDermott, Van Horn and Hansen \cite{mhh88}.
But we have shown in Sec. IIIA that the elastic-response assumption
is not valid during high accretion-rate intervals which necessarily
produce nuclear rearrangement within the solid structure.
In Sec. IIB and IIIB, we have shown that the basic boundary condition
in the accretion problem is that the pressure $P_{c}$ at the crust-core
boundary is a constant.  In the Cowling approximation, and with neglect
of the effects of rotation and thermal fluctuations, the boundary
surface is a sphere whose radius $R_{c}$, for most equations of state,
decreases slowly as the mass
of the star increases.  There is no impediment to the maintenance of
this condition by the rapid transfer of matter between the solid and
liquid phases through electromagnetic and strong interactions, and
there is no
immediate local dependence on weak-interactions transition rates.
This condition leads to the very simple one-dimensional solutions
Eq. (7) and (8) of Sec. IIIB which appear hydrostatic in nature.
Nonetheless, the stability of the system obviously depends on the
shear modulus of the solid.
  
As a source of gravitational radiation, it appears from our analysis
that the deformation
produced by angle-dependence of the {\it rp} process end-point is not
quantifiable and may be no more important than that derived from the
Maxwell tensor in the solid
crust. The two sources are not distinguishable unless there is some
{\it a priori} information about the internal field.  The uncertainties
exposed by our analysis seem unlikely to be removed in the near future
and indicate that it will not be possible to decide with any confidence
whether angle-dependent composition or  
the magnetic structure and superconductivity of the core \cite{cut02}
is the more probable origin of any periodic signals seen in
future gravitational wave experiments \cite{abb04}. 

\bibliography{apssamp}% Produces the bibliography via BibTeX.

\end{document}